\documentclass[twocolumn]{aastex701}
\usepackage{amsmath}

\newcommand{\spi}{{\it Spitzer}}

\newcommand{\oii}{\hbox{[O$\,${\scriptsize II}]}}

\newcommand{\mgii}{\hbox{Mg$\,${\scriptsize II}}}

\newcommand{\neii}{\hbox{[Ne$\,${\scriptsize II}]}}
\newcommand{\neiii}{\hbox{[Ne$\,${\scriptsize III}]}}

\newcommand{\kms}{km\,s$^{-1}$} 
\newcommand{\msun}{M$_{\odot}$}

\newcommand{\ergs}{erg s$^{-1}$}

\newcommand\jwst{\emph{JWST}}
\newcommand\hst{\emph{HST}}
\newcommand\chandra{\emph{Chandra}}
\newcommand\makani{\emph{Makani}}
\newcommand\stpsf{\emph{STPSF}}

\newcommand\photutils{\texttt{photutils}}
\newcommand\psfphotometry{\texttt{PSFPhotometry}}

\shorttitle{Warm Dust in the Circumgalactic Medium of \makani}
\shortauthors{Veilleux al.}
\graphicspath{{./}{figures/}}

\begin{document}

\title{\jwst\ Discovery of Warm Dust in the Circumgalactic Medium of the \makani\ Galaxy}

\author[0000-0002-3158-6820]{Sylvain Veilleux}
\affiliation{Department of Astronomy, University of Maryland, College Park, MD 20742, USA}
\affiliation{Joint Space-Science Institute, University of Maryland, College Park, MD 20742, USA}
\email{veilleux@astro.umd.edu}

\author{Steven D. Shockley}
\affiliation{Department of Astronomy, University of Maryland, College Park, MD 20742, USA}
\email{sshockle@terpmail.umd.edu}

\author[0000-0001-8485-0325]{Marcio Mel\'endez}
\affiliation{Space Telescope Science Institute, Baltimore, MD 21218, USA}
\email{melendez@stsci.edu}

\author[0000-0002-1608-7564]{David S. N. Rupke}
\affiliation{Department of Physics, Rhodes College, Memphis, TN 38112, USA}
\affiliation{Zentrum f\"ur Astronomie der Universit\"at Heidelberg, Astronomisches Rechen-Institut, M\"unchhofstr 12-14, D-69120 Heidelberg, Germany}
\email{rupked@rhodes.edu}

\author[0000-0002-2583-5894]{Alison L. Coil}
\affiliation{Department of Astronomy and Astrophysics, University of California, La Jolla, CA 92092, USA}
\email{acoil@ucsd.edu}

\author{Aleksandar M. Diamond-Stanic}
\affiliation{Department of Physics and Astronony, Bates College, Lewiston, ME 04240, USA}
\email{adiamond@bates.edu}

\author[0000-0003-4964-4635]{James E. Geach}
\affiliation{Centre for Astrophysics Research, Department of Physics, Astronomy and Mathematics, University of Hertfordshire, Hatfield, UK AL10 9AB}
\email{j.geach@herts.ac.uk}

\author[0000-0003-1468-9526]{Ryan C. Hickox}
\affiliation{Department of Physics and Astronomy, Dartmouth College, Hanover, NH 03755, USA}
\email{ryan.c.hickox@dartmouth.edu}

\author[0000-0002-2733-4559]{John Moustakas}
\affiliation{Department of Physics and Astronomy, Siena College, 515 Loudon Road, Loudonville, NY 12110, USA}
\email{jmoustakas@siena.edu}

\author[0000-0001-5851-1856]{Gregory H. Rudnick}
\affiliation{Department of Physics and Astronomy, University of Kansas, Lawrence, KS 66045, USA}
\email{grudnick@ku.edu}

\author[0000-0003-1771-5531]{Paul H. Sell}
\affiliation{Department of Astronomy, University of Florida, Gainesville, FL 32611, USA}
\email{psell@ufl.edu}

\author[0000-0003-3097-5178]{Christy A. Tremonti}
\affiliation{Department of Astronomy, University of Wisconsin-Maddison, Maddison, WI 53706, USA}
\email{tremonti@astro.wisc.edu}

\author{Hojoon Cha}
\affiliation{Department of Astronomy, University of Wisconsin-Maddison, Maddison, WI 53706, USA}
\email{hcha9@wisc.ed}



\begin{abstract}
We report the detection of near- and mid-infrared emission from polycyclic aromatic hydrocarbons (PAHs) out to $\sim$ 35 kpc in the \makani\ Galaxy, a compact massive galaxy with a record-breaking 100-kpc scale starburst-driven wind at redshift $z$ = 0.459. The NIRCam and MIRI observations with \jwst\ take advantage of a coincidental match between the PAH spectral features at 3.3, 7.7, and (11.3 + 12.2) \micron\ in \makani\ and the bandpasses of the MIRI and NIRCam filters. The warm dust is not only detected in the cool-gas tracers of the galactic wind associated with the more recent (7 Myr) starburst episode, but also in the outer warm-ionized gas wind produced by the older (0.4 Gyr) episode.  The presence of PAHs in the outer wind indicates that the PAHs have survived the long ($R/v$ $\sim$ 10$^8$ yrs) journey to the halo despite the harsh environment of the galactic wind. The measured F1800W/F1130W flux ratios in the unresolved nucleus, inner halo ($R$ = 10 $-$ 20 kpc), and outer halo ($R$ = 20 $-$ 35 kpc), tracers of the PAH (11.3 + 12.2)/7.7 ratios, indicate decreasing starlight intensity incident on the PAHs, decreasing PAH sizes, and increasing PAH ionization fractions with increasing distance from the nucleus. These data provide the strongest evidence to date that the ejected dust of galactic winds survives the long journey to the CGM, but is eroded along the way.
\end{abstract}



\section{Introduction} 
\label{sec:intro}

The amount of dust outside galaxies, inferred from reddening measurements of background quasars and galaxies by foreground galaxy halos \citep[][]{Menard2010, Peek2015, McCleary2025}, is comparable to that within galaxies, but the origin of this dust is uncertain \citep[e.g.,][and references therein]{Tumlinson2017, Faucher2023}. Sophisticated cosmological galaxy formation simulations \citep[e.g.,][] {Vogelsberger2014, Schaye2015, Dubois2016, Choi2018, Biernacki2018, Brennan2018, Hopkins2018, Dave2019, Nelson2019, Peeples2019, Hafen2019, Narayanan2023} 
suggest that galactic winds, driven by stellar or supermassive black hole (SMBH) processes, are the primary source of the enriched circumgalactic medium (CGM), but the direct detection of dust in a galactic wind on the relevant CGM scale remains elusive \citep[e.g.,][and references therein]{Veilleux2020}.  

The recently discovered 100-kpc scale wind in the \makani\ Galaxy \citep[SDSS J211824.06+001729.4;][]{Rupke2019}, a massive ($M_*$ = 10$^{11.1}$ \msun) but compact ($r_e$ = 2.3 kpc) post-starburst galaxy at $z$ = 0.459 \citep{Sell2014}, is an excellent target for testing this idea. The cooler neutral-atomic and molecular gas phases in this wind coexist with the warm ionized gas to distances of $\sim$ 20 kpc but apparently not beyond. This provides tantalizing evidence that we are witnessing, for the first time on the CGM scale, the dissolution of the outflowing cool clouds into the warm ionized phase as theoretically predicted \citep[e.g.,][]{Scannapieco2015, Ferrara2016, Decataldo2017}.  The lack of significant reddening in the warm ionized gas beyond the cool molecular/neutral-atomic gas \citep[based on the long-slit measurements of the H$\alpha$/H$\beta$ emission line ratio from][]{Rupke2023} further suggests that dust does not survive the journey beyond $\sim$ 25 kpc.

The exquisite infrared sensitivity and image quality of \jwst\  provide new tools to examine this issue. Indeed several MIRI and NIRCam filters \citep[][]{Gardner2023, Rigby2023} are specifically designed to capture key spectral features from warm dust (PAH 3.3, 7.7, 11.2, and 17.0) and molecular gas (H$_2$ 0-0 S(1) 17.03). Multi-band NIRCam and MIRI images of the classic starburst galaxy M82 have beautifully demonstrated the ability of \jwst\ to map PAHs in nearby galactic wind systems \citep[][]{Bolatto2024, Fisher2025}. However, these PAH features are redshifted out of the filter bandpasses in galaxies with $z \ga 0.03$. The \makani\ Galaxy is a remarkable exception: at $z = 0.459$, the bandpasses of seven MIRI and NIRCam filters coincide with redshifted Pa$\beta$, PAH 3.3, PAH 7.7, (PAH 11.3 + 12.7), (PAH 17.0 + H$_2$ S(1) 17.03), and two largely featureless continuum windows (Figure \ref{fig:filter_throughput_makani}). Observations through these filters therefore offer a unique window on the warm dust, H$_2$, and ionized gas contents across the entire galactic wind of {\em Makani}. 

This paper reports the results of our analysis of the multi-filter MIRI and NIRCam data on \makani\ obtained by our group on 2023-06-30 (PID GO-1865, PI Veilleux). These data are publicly available on the Mikulski Archive for Space Telescopes (MAST) at the Space Telescope Science Institute. The specific observations analyzed can be accessed via \dataset[doi: 10.17909/1r37-nb46]{https://doi.org/10.17909/1r37-nb46}.
Section \ref{sec:obs} briefly describes the observing strategy and MIRI and NIRCam observations. The results from our analysis are presented in Section \ref{sec:results} and the implications of these results on the issues of dust survival and evolution are discussed in Section \ref{sec:discussion}. The conclusions are summarized in Section \ref{sec:conclusions}. 
Throughout this paper, we take as systemic the stellar redshift of \makani, $z=0.4590$ \citep{Rupke2019} and assume a flat $\Lambda$ cosmology with $\Omega_m=0.315$ and $H_0=67.4$~\kms~Mpc$^{-1}$ \citep{Planck2018}, resulting in a projected physical scale of 6.02 kpc arcsec$^{-1}$ at the distance of \makani.

\section{Observations} 
\label{sec:obs}

\subsection{Observing Strategy}

\begin{figure}
    \centering
    \includegraphics[width=0.45\textwidth]{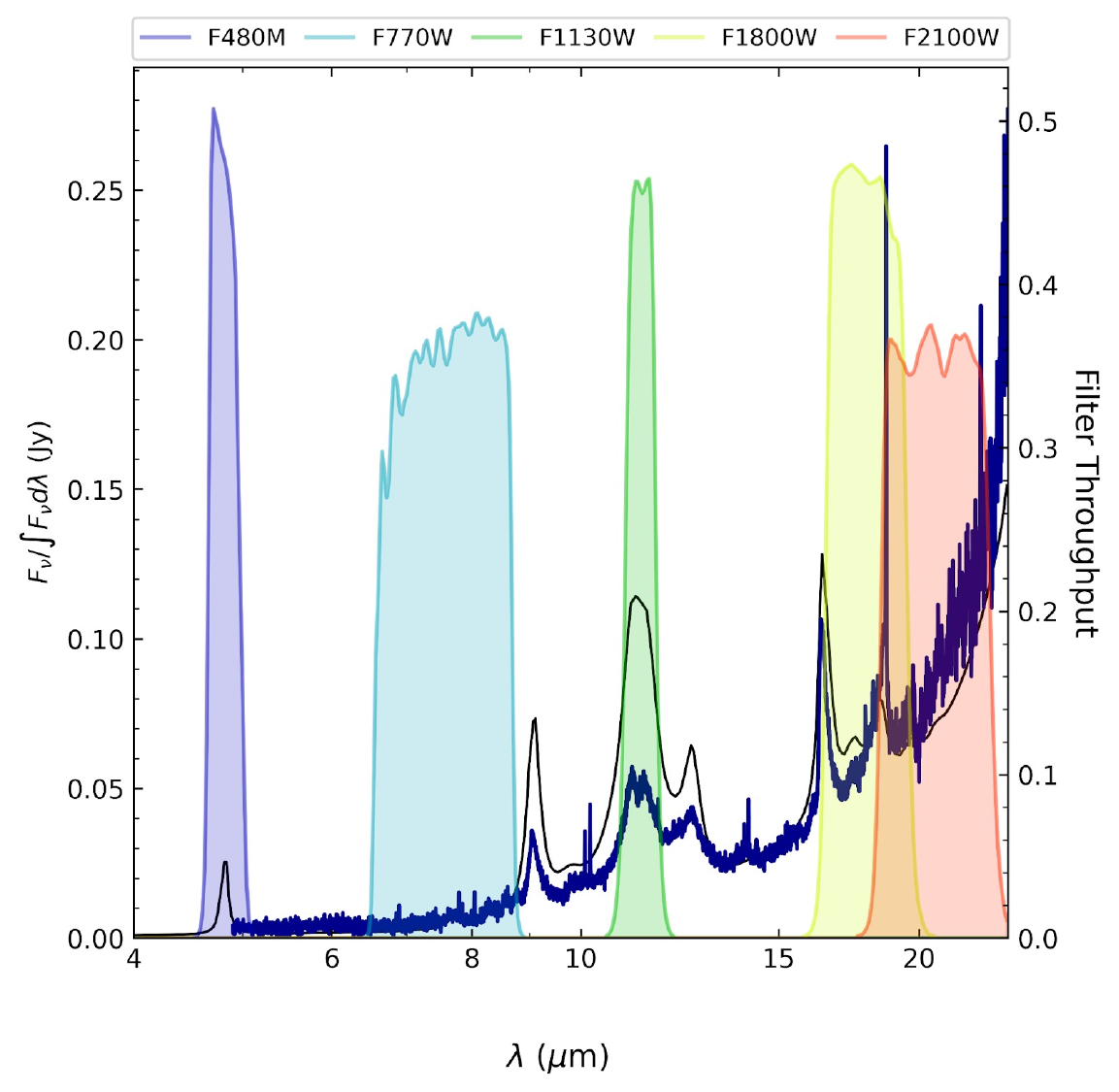}
    \caption{The transmission curves of both the NIRCam and MIRI filters are overlaid on the nuclear spectrum of \makani\ (from Erena et al.\ 2025, in prep.; thick line) and one of the model spectra of \citet[][thin line]{Draine2021}. The wavelength is in the observer's frame. The left axis shows the normalized spectral intensity and the right axis indicates the filter throughput. Several filters are centered on distinct PAH features while others cover the featureless continuum. Non-PAH features, such as fine-structure atomic lines, contaminate the fluxes within some filters, but their overall contributions can safely be ignored. See Col.\ (3) in Table \ref{tab:observation_summary} and Section \ref{sec:contamination} for more details.}
    \label{fig:filter_throughput_makani}
\end{figure}

\makani\ is at a very convenient redshift: the bandpass of F480M is centered on redshifted PAH 3.3 \micron, F1130W on redshifted PAH 7.7 \micron, F1800W on redshifted PAH (11.3 + 12.2) \micron, and F2550W on redshifted H$_2$ 17.03 \micron, while F770W and F2100W cover the largely featureless continuum emission at rest-frame 4.5 $-$ 6 \micron\ and 13 $-$ 16 \micron, respectively (Table \ref{tab:observation_summary}, Figure \ref{fig:filter_throughput_makani}). Moreover, F187N, which comes for free while observing with F480M, also happens to be centered on redshifted Pa$\beta$. 
The flux measurements from F770W and F2100W constrain the strength and slope of the continuum emission without PAH contamination. They can therefore be used in principle as ``off-band" filters to estimate the continuum-subtracted fluxes of the PAH 3.3, 7.7, (11.3 + 12.2), and H$_2$ 17.03 features. Some weaker features contaminate our measurements (col.\ 3 in Table \ref{tab:observation_summary}), but their contributions are negligibly small (this is discussed in Section \ref{sec:contamination} below).

\begin{deluxetable*}{l c c c}
 \tablecaption{NIRCam and MIRI Observations\label{tab:observation_summary}}
 \tablehead{
 \colhead{Filters} & \colhead{Main} & \colhead{Weaker} & \colhead{$t_{\rm exp}$}\\
\colhead{} & \colhead{Diagnostics} & \colhead{Feature(s)}& \colhead{(hr)}\\
 \colhead{(1)} & \colhead{(2)} & \colhead{(3)} & \colhead{(4)}}
 \startdata
N-F187N & Pa$\beta$ 1.28 & --- & 0.90 \\ 
N-F480M & PAH 3.3 & --- & 0.90 \\ 
M-F770W & Continuum (5 $\mu$m) & --- & 0.56 \\ 
M-F1130W & PAH 7.7 & --- & 0.44 \\
M-F1800W & PAH$\,$(11.3 + 12.2) & [Ne II]$\,$12.8 & 0.47 \\
M-F2100W & Continuum (14 $\mu$m) & [Ne II]$\,$12.8 + [Ne~III]$\,$15.5 & 0.78 \\
M-F2550W & H$_2$ 17.03 & PAH 17.0 & 1.00 \\
 \enddata
 \tablecomments{Meaning of the columns: (1) Name of the NIRCam (N) or MIRI (M) filters. (2) Main diagnostics in the filter bandpass. (3) Weaker feature(s) in the filter bandpass. (4) On-target exposure in hrs (excl.\ overheads). } 
\end{deluxetable*}

\subsection{MIRI Observations}

The MIRI exposures were broken into short segments to avoid saturation on the central galaxy core. For all observations, the recommended default readout mode for imaging, FAST, and 4-point extended-source dither pattern were used. The F770W, F1130W, and F1800W observations were split into 60 groups, 1 integration, and 12 exposures, 35 groups, 1 integration, and 16 exposures, and 30 groups, 1 integration, and 20 exposures, respectively, keeping the integration time at each dither position below 1,000 seconds. For the long-wavelength F2100W and F2550W observations, the maximum dwell time at each dither position was kept well below 8 minutes using 28 groups, 1 integration, and 36 exposures, and 20 groups, 1 integration, and 64 exposures, respectively. The MIRI FULL array field of view (FOV; 74\arcsec\ $\times$ 114\arcsec) is large enough to cover the full extent of {\em Makani} and provides source-free sky beyond the galactic wind, so separate background observations were not requested.

\subsection{NIRCam Observations}

A similar strategy was used for the NIRCam F480M observations. To reach a good signal-to-noise ratio (S/N) in the wind, avoid saturation on the galaxy, and reduce overheads and data volume, an on-target exposure time of 1 hr was split into 10 groups, 1 int., and 20 exp. using the MEDIUM8 readout pattern and subarray SUB400P. This number of groups mitigates cosmic rays for all pixels and maximizes S/N while keeping the integration times per exposure well below 1000 seconds, as recommended. The recommended larger subpixel STANDARD dithers were used to better mitigate bad pixels in the case where no primary dithers are performed. 
Separate background observations were not requested since the FOV of the SUB400P subarray of the long-wavelength side of NIRCam (25\farcs0 $\times$ 25\farcs0) is large enough to cover the full extent of {\em Makani} and provide source-free sky beyond the wind. The FOV on the short-wavelength side of NIRCam (12\farcs4 $\times$ 12\farcs4) covers most of the [O~II] galactic wind and some source-free sky along the short axis of the wind.

\begin{figure*}[htb!]
    \centering
    \includegraphics[width=0.95\textwidth]{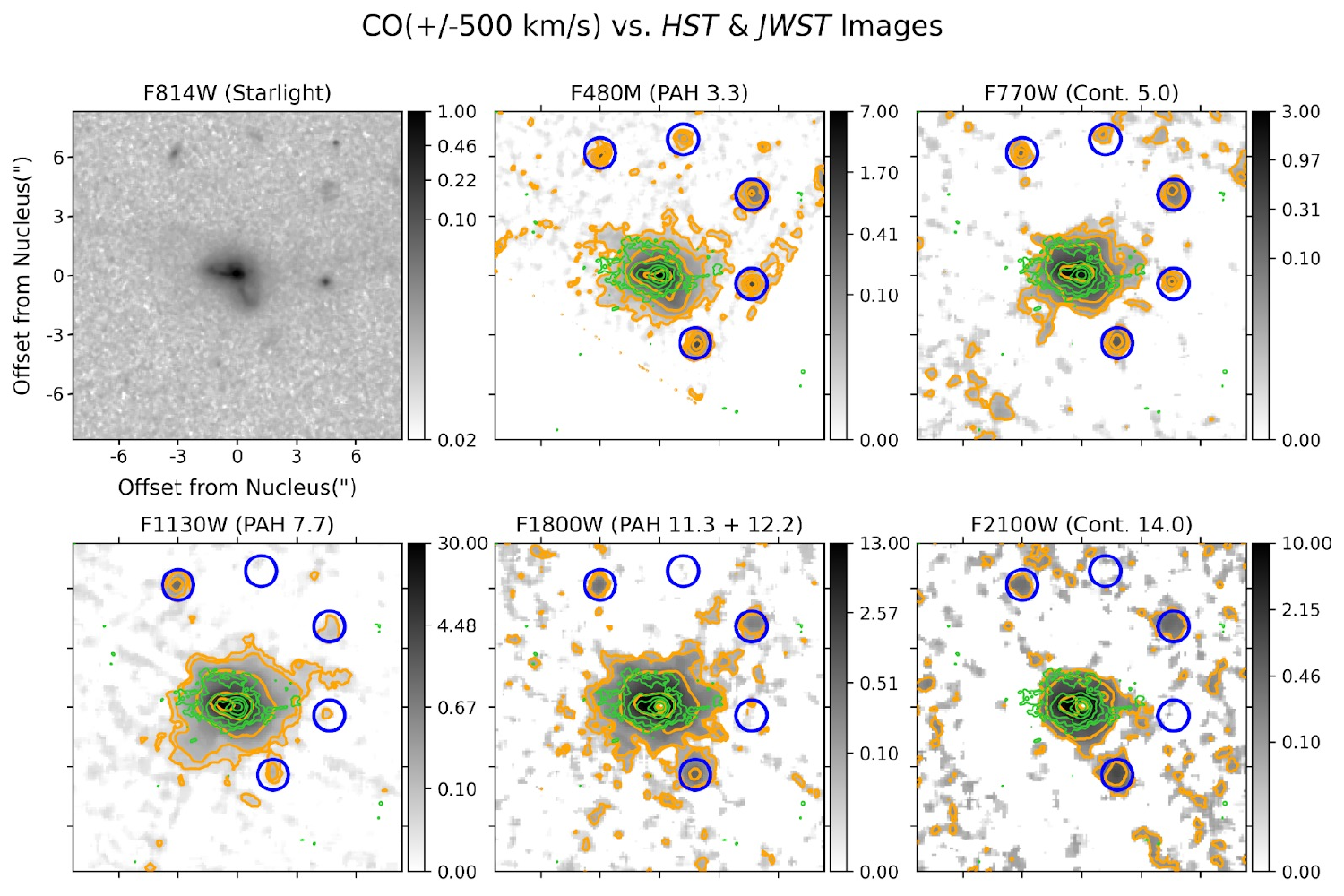}
    \caption{The top-left panel shows the stellar emission of \makani\ at rest-frame $\sim$ 5600 \AA\ mapped in the \hst\ F814W filter. The grey scale for that panel is in counts sec$^{-1}$ pixel$^{-1}$. The other panels show the flux maps of the dust emission (orange contours) in the \jwst\ NIRCam F480M and MIRI F770W, F1130W, F1800W, and F2100W filters after PSF subtraction with \stpsf. The field of view of each panel is approximately 100 $\times$ 100 kpc. The grey scale for the \jwst\ data is in units of MJy sr$^{-1}$. The orange contour levels are 0.01, 0.03, 0.1, 0.3, and 3.0 MJy sr$^{-1}$ for F480M, 0.01, 0.03, 0.1, and 1.0 MJy sr$^{-1}$ for F770W, 0.05, 0.1, 0.5, 5.0, and 15.0 MJy sr$^{-1}$ for F1130W, 0.05, 0.1, 0.5, and 5.0 MJy sr$^{-1}$ for F1800W, and 0.1, 0.5, and 5.0 MJy sr$^{-1}$ for F2550W. The CO (2-1) line emission within [$-$500, $+$500] \kms\ from \citet{Rupke2019} is shown for comparison as green contours where the contour levels are 0.09, 0.14, 0.28, 0.5, 0.8, and 1.2 mJy beam$^{-1}$.  The blue circles identify possible galaxies in the field.}
    \label{fig:jwst-co}
\end{figure*}

\begin{figure*}[htb!]
    \centering
    \includegraphics[width=0.95\textwidth]{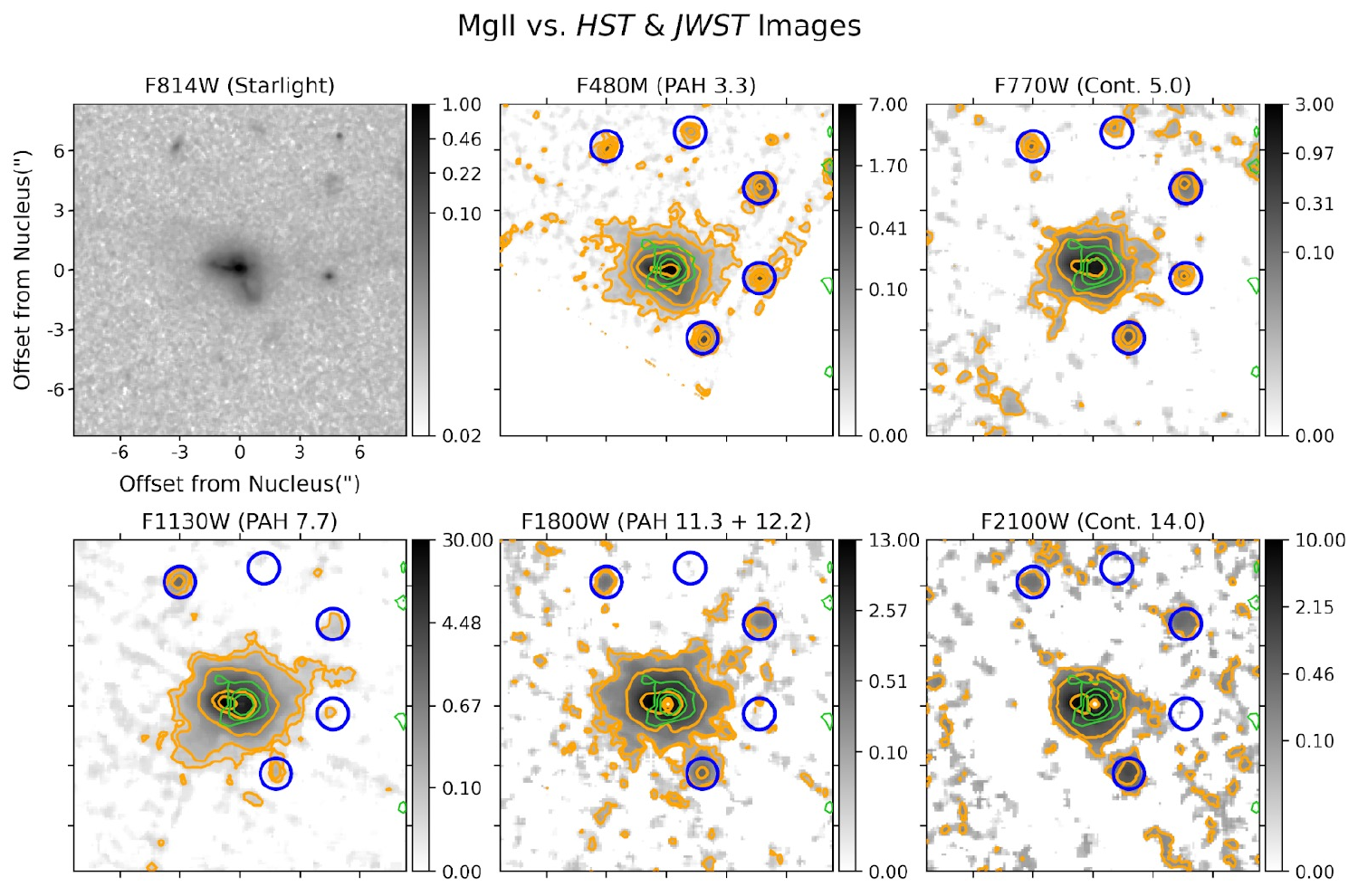}
    \caption{Same as Figure \ref{fig:jwst-co} but here the green contours show the \mgii\ 2800 \AA\ line emission instead of CO (2-1). The green contour levels are 0.15, 0.29, and 0.59 $\times$ 10$^{-16}$ \ergs\ cm$^{-2}$ arcsec$^{-2}$.}
    \label{fig:jwst-mgii}
\end{figure*}

\begin{figure*}[htb!]
    \centering
    \includegraphics[width=0.95\textwidth]{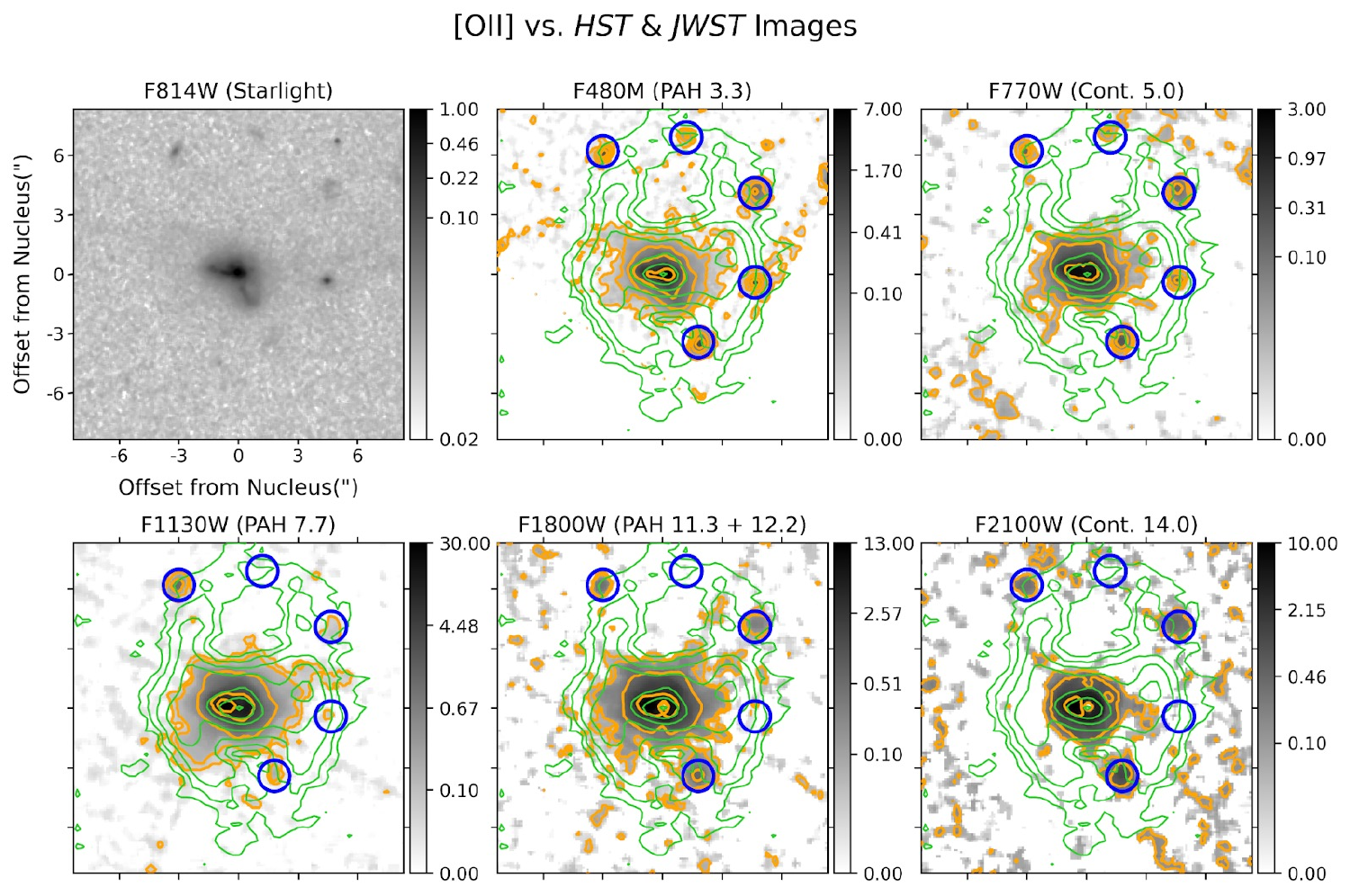}
    \caption{Same as Figure \ref{fig:jwst-co} but here the green contours show the \oii\ 3727 \AA\ line emission instead of CO (2-1). The green contour levels are 0.037, 0.074, 0.15, 0.29, 0.59, 1.2, 2.4 $\times$ 10$^{-16}$ \ergs\ cm$^{-2}$ arcsec$^{-2}$.}
    \label{fig:jwst-oii}
\end{figure*}

\section{Results}
\label{sec:results}

\subsection{Data Reduction and Analysis}
\label{sec:data-reduction-analysis}

All data used for this analysis were processed using \jwst\ Calibration pipeline version 1.10.1. We used the dither-combined stage 3 products (i2d fits) from the Mikulski Archive for Space Telescopes (MAST).

We are interested in the faint extended emission outside of the bright central starburst of \makani. Therefore, we need to carefully remove the bright central point spread function (PSF). 
To model the \jwst\ PSF, we employed \stpsf\ \citep[version 2.0.0;][]{Perrin2014}, a Python package that leverages in-flight wavefront measurements to simulate the observed PSFs taking into account detector pixel scales, rotations, filter profiles and detector effects. Note that \stpsf\ incorporates simplified models for detector effects like inter-pixel capacitance, charge diffusion, and the MIRI cruciform artifact. While the current model accounts for field-dependence in the cruciform artifact, discrepancies between predicted and observed flux persist.

We modeled the optimal PSF for our June 30, 2023 \makani\ observations using the closest-in-time optical path difference map. These wavefront measurements, regularly available on MAST, enable \stpsf\ to generate time-dependent PSFs that accurately reflect the telescope's optical performance at the time of the science observations. Subsequently, we utilized the \psfphotometry\ module from the \photutils\ Python package to create residual images. This involved fitting and subtracting our \stpsf\ PSF model from the \makani\ observations at each wavelength.

\subsection{Continuum and Emission Line Contamination}
\label{sec:contamination}

The stellar continuum from the galaxy is negligible beyond $r$ $\simeq$ 1\arcsec\ (6 kpc) from the center of this compact galaxy \citep[$r_{1/2}$ = 2.3 kpc;][]{Sell2014}. However, some fine-structure lines may contaminate our measurements (col.\ 3 in Table \ref{tab:observation_summary}, Figure \ref{fig:filter_throughput_makani}):  \neii\ 12.8 in F1800W and  both \neii\ 12.8 and \neiii\ 15.5 in F2100W (we ignore the PAH 17.0 + H$_2$ blend in F2550W since no extended emission is detected in this filter). 
This flux correction is only $\sim$ 1\% in the nuclear spectrum of \makani\ (Figure \ref{fig:filter_throughput_makani}) and typically less than 5\% of the PAH (11.3 + 12.2) flux in the wind of M82 \citep[][]{Beirao2015, Bolatto2024}. Similarly, the \neiii\ contribution to the rest-frame 14 $\mu$m continuum emission in the F2100W filter is less than 1\% in the nuclear spectrum of \makani\ and $<$5\% assuming \neiii\ 15.5 / \neii\ 12.8 $\simeq$ 1/3 \citep[e.g., M82 wind:][]{Beirao2015}. Both \neii\ and \neiii\ therefore contribute negligibly to the measured fluxes in filters F1800W and F2100W; they are ignored in the remainder of our discussion.  

\subsection{Distribution of the Warm Dust Relative to the Cool and Warm Gas Phases}
\label{sec:distributions}

The results from the PSF subtraction with \stpsf\ are shown in Figures \ref{fig:jwst-co} $-$ \ref{fig:jwst-oii}. 
To help assess the full extent of the extended emission associated with \makani, we directly compare the PSF-subtracted MIRI and NIRCam images of \makani\ with 
the \hst\ F814W (stellar continuum near $\sim$6000 \AA\ in the rest-frame), ALMA CO (2-1), KCWI \mgii\ 2800 \AA, and KCWI \oii\ 3727 \AA\ flux maps from \citet{Rupke2019}. The blue circles in these figures identify possible galaxies in the field. Figures \ref{fig:jwst-co} and \ref{fig:jwst-mgii} show that warm dust is clearly detected beyond the inner cold molecular gas traced by CO (2$-$1) and the inner cool neutral-atomic gas traced by \mgii\ (IP = 8 eV), respectively. On the other hand, Figure \ref{fig:jwst-oii} shows that the mid-infrared emission is less extended than the warm ionized gas traced by \oii. 
Not shown in these figures, the redshifted PAH 17.0 + H$_2$ 17.03 blend and underlying continuum emission probed by the F2550W filter remains undetected beyond $\sim$ 1\arcsec\ due to the lower-than-expected sensitivity of MIRI at 25 $\mu$m and the contaminating emission from the central unresolved galaxy, which is not only more extended at these long wavelengths due to the wavelength dependence of the diffraction pattern but also brighter (Figure \ref{fig:filter_throughput_makani}). Similarly, the F187N image, which came for free with the F480M observations, is too shallow to reveal extended redshifted Pa$\beta$ line emission beyond $\sim$ 1\arcsec. The F187N and F2550W observations are not discussed any further in this paper.

\begin{figure}[htb!]
    \centering
    \includegraphics[width=0.90\textwidth]{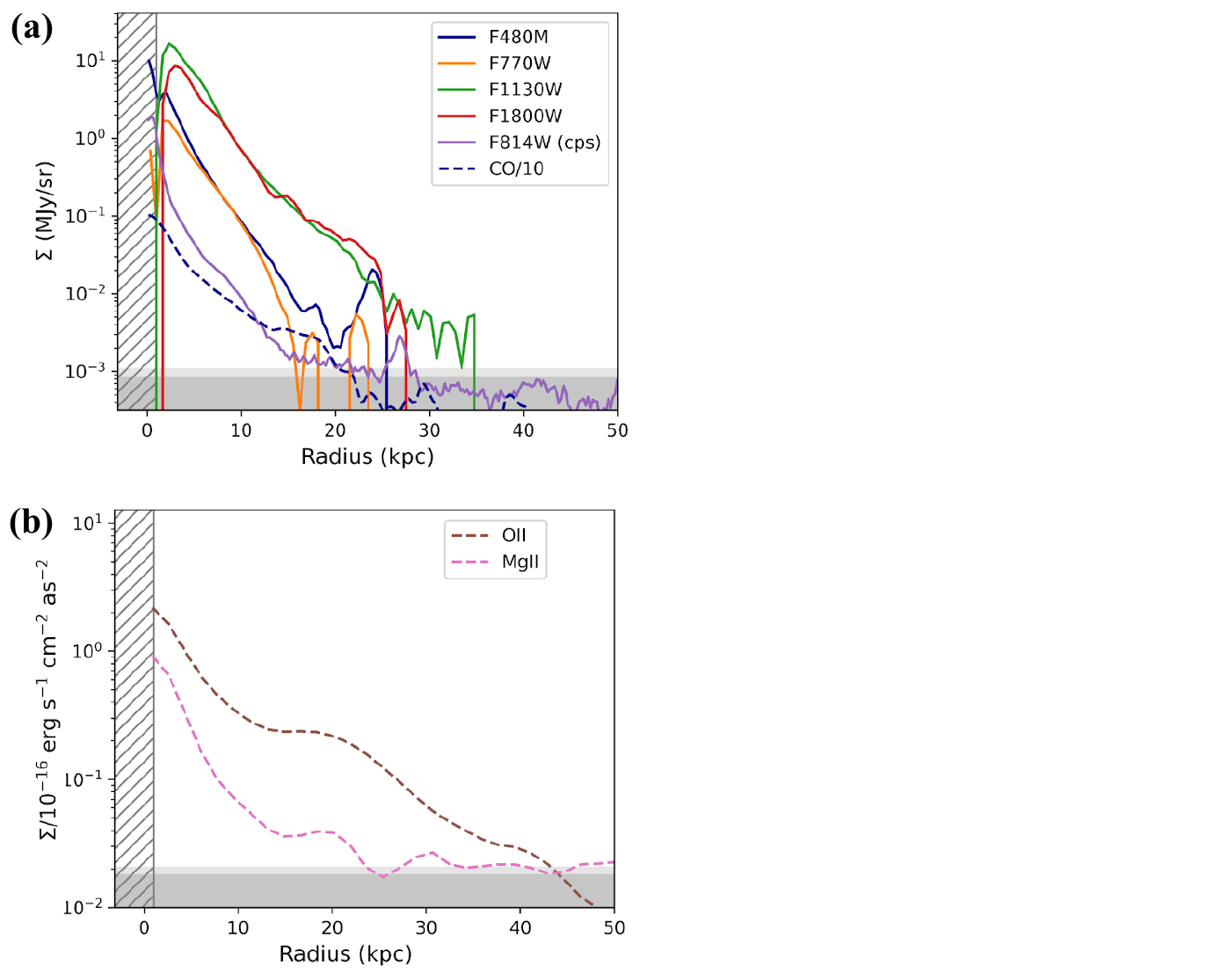}
    \caption{(a) Radial surface brightness profiles in  the \jwst\ filters (after PSF subtraction with \stpsf), the \hst\ F814W filter, which probes starlight near $\sim$ 5600 \AA, and the line emission in CO (2-1) (within [$-$500, $+$500] \kms). (b) Same as (a) but for \oii\ 3727 \AA\ and \mgii\ 2800 \AA. Measurements in the vertical hatched region at radii smaller than $\sim$ 2 kpc are not shown since they are affected by the PSF subtraction. The horizontal grey areas in (a) and (b) are the detection thresholds of the CO (2-1) and \mgii\ surface brightness measurements, respectively. }
    \label{fig:radial}
\end{figure}

In Figure \ref{fig:radial}, we compare the azimuthally averaged surface brightness radial profiles of \oii, \mgii, and CO 2$-$1 from \citet{Rupke2019, Rupke2023} with the PAH emission at 3.3, 7.7, and (11.3 + 12.2) derived from the NIRCam and MIRI PSF-subtracted images shown in Figures \ref{fig:jwst-co} $-$ \ref{fig:jwst-oii}.  This figure highlights the detection of extended warm dust emission in the images of F1130W (rest-frame PAH 7.7 $\mu$m) and F1800W (rest-frame PAH 11.3 + 12.2 $\mu$m), and to a lesser extent in the images of F480M (rest-frame PAH 3.3 $\mu$m) and F770W (rest-frame continuum 5 $\mu$m), well beyond the CO 2-1 and \mgii\ line emission which is not detected beyond $\sim$ 20 kpc. Note, however, that the F1130W/CO (2$-$1), F1800W/CO (2$-$1), F1130W/\mgii, and F1800W/\mgii\ flux ratios remain relatively constant out to 20 kpc. We discuss the physical implications of these results in Section \ref{sec:discussion}.

\begin{figure*}[htb!]
    \centering
    \includegraphics[width=0.95\textwidth]{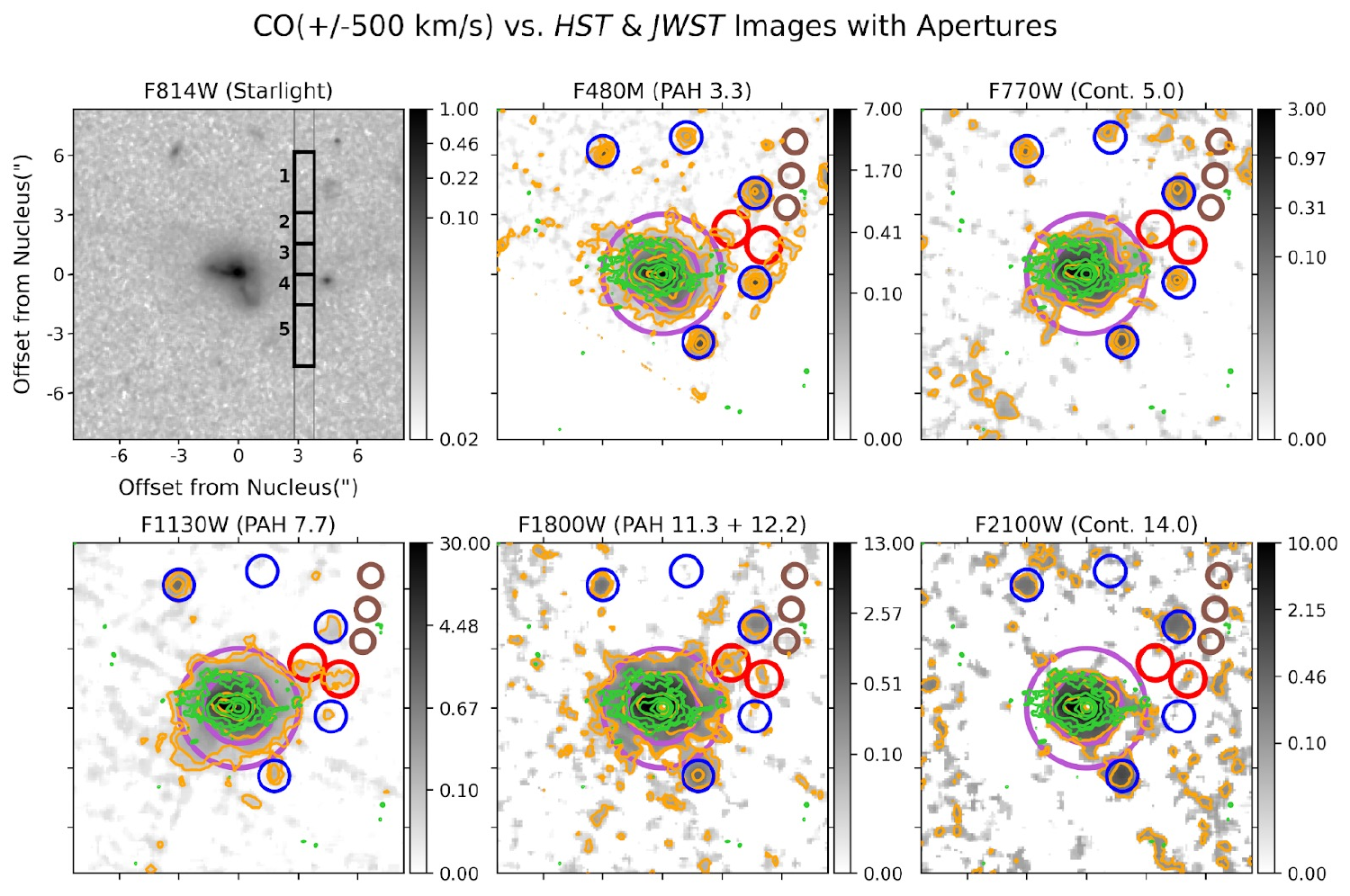}
    \caption{Same as Figure \ref{fig:jwst-co} but now showing the apertures used to calculate the strength of the dust emission in the inner halo (purple annulus centered on the nucleus with inner and outer radii of 1\farcs78 and 3\arcsec, respectively) and outer halo (CGM-E and CGM-W; eastern and western red circles, respectively).  The flux measurements are listed in Table \ref{tab:fluxes}. The background level was assessed within the three circular apertures in the upper right corner of each panel, far from any galaxy wind emission. In the upper left panel, we also show the locations of some of the extraction apertures along the Keck/ESI long-slit spectrum used by \citet{Rupke2023}.}
    \label{fig:apertures}
\end{figure*}

In addition, clear asymmetries in the mid-infrared emission are visible in Figures \ref{fig:jwst-co} $-$ \ref{fig:jwst-oii}, but are not captured in the azimuthally-averaged radial profiles shown in Figure \ref{fig:radial}. In particular, a cloud complex is visible in the outer halo of the PSF-subtracted F1130W and F1800W images, extending out to 30 $-$ 35 kpc NW of the galaxy nucleus. The (upper limits on the) flux densities measured within two circular apertures that encompass these clouds are listed in Table \ref{tab:fluxes} (the footprint of these apertures are shown in red in Figure \ref{fig:apertures}). For comparison, we also measure the fluxes in the nucleus
and the inner halo, defined as the annular region centered on the nucleus with inner and outer radii of 1\farcs78 (10.5 kpc) and 3\arcsec (17.6 kpc), respectively, shown in purple in Figure \ref{fig:apertures}. The variations in the flux ratios within the various apertures are discussed in Section \ref{sec:discussion}. 

\section{Discussion}
\label{sec:discussion}

Here we discuss the implications of the \jwst\ results on the issue of dust survival
in the circumgalactic environment (Section \ref{subsec:survival}) and how dust evolves as it is carried away by the wind to these large distances (Section \ref{subsec:evolution}).

\subsection{Warm Dust in the Outer Wind of \makani}
\label{subsec:survival}

The spatially-resolved \oii\ kinematics derived from the KCWI data cube in \citet{Rupke2019} cleanly separate the wind into an outer region ($r$ = 20 $-$ 50 kpc) made of warm ionized gas with maximum blueshifted velocities ($v_{98\%}$) between $-$100 and $-$700 \kms\ and an inner region ($r$ $\la$ 20 kpc) where the warm ionized gas traced by \oii\ has a high velocity tail to $-$2100 \kms. \citet{Rupke2019} have argued that these two wind components were produced by the two distinct bursts of star formation detected in the central core, Episode I (0.4 Gyr ago) and Episode II (7 Myr ago). The Episode II ionized outflow is accompanied by two other gas phases: high-velocity (up to $\pm$ 1500 km s$^{-1}$) cold molecular gas traced by mm-wave CO(2-1) line emission and a cool neutral-atomic gas in the velocity range $\pm$ 500 km s$^{-1}$ traced by resonant line emission from Mg~II 2796, 2803 \AA\ (IP = 8 eV).
 
The extended PAH and infrared continuum emission shown in Figures \ref{fig:jwst-co} $-$ \ref{fig:jwst-oii} and Figure \ref{fig:radial} traces the distribution of the warm dust across {\em both} the inner and outer winds of \makani. The presence of dust in the outer warm ionized wind is unexpected since the optical reddening due to dust, $E(B-V)$, derived by \citet{Rupke2023} from the Keck/ESI long-slit measurements of the H$\alpha$/H$\beta$ ratios assuming Case B recombination, decreases with radius from a peak value of $\sim$ 1 mag.\ at $R$ = 10 kpc to $\sim$ 0 mag.\ at $R >$ 25 kpc.  Non-zero extinction is measured at the base of this cloud complex (e.g., $E(B-V)$ = 0.42 $\pm$ 0.06 mag.\ at $R \sim$ 21 kpc in aperture \#3 defined\footnote{Note that the aperture definition here is different from that of \citet[][]{Rupke2023}.} in the upper left panel of Fig.\ \ref{fig:apertures}),
but not beyond (e.g., in aperture \#2 at $R \sim$ 25.5 kpc). 
We note, however, that the long-slit measurements of \citet{Rupke2023} do not fully sample the NW warm-dust clouds at $R \ga$ 25 kpc (e.g., Fig.\ \ref{fig:apertures}), so it is quite possible that the slit missed some of the dust.

The presence of dust in the outer Episode I warm ionized nebula speaks to the durability of dust in the galactic wind environment.  The dynamical time of the outflowing dusty gas to reach distances of $\sim$ 30 kpc is $t_{\rm dyn} \sim$ $R/V \sim$ 0.1 Gyr, assuming that the dust is entrained at the same velocity as the gas at an average velocity of 300 \kms. This timescale may be compared with the dust erosion/destruction timescale. Several processes may be at work in the wind: e.g. collisions with other grains (shattering), sputtering due to collisions with ions, sublimation or evaporation, explosion due to UV radiation, alteration of grain material by cosmic rays and X-rays. The thermal sputtering time of a dust grain of size $a$ in a hot plasma of density $\rho$ and temperature $T$ is approximately given by  
\begin{equation}
\begin{aligned}
        t_{\rm sp} \approx\ 0.17~{\rm Gyr} & \left(\frac{a}{0.1~\mu m}\right) \left(\frac{10^{-27} {\rm g~cm}^{-3}}{\rho}\right) \\ 
    & \times \left[\left(\frac{10^{6.3}~{\rm K}}{T}\right)^\omega + 1\right], 
\end{aligned}
\end{equation}
where $\omega$ = 2.5 
\citep{Tsai1995, Hirashita2015, Mckinnon2017, Richie2024}.

Using the optical line flux ratios [O~II] 3729 \AA/3726 \AA\ and [S~II] 6716 \AA/6731 \AA, \citet{Rupke2023} derive an upper limit on the electron density $n_{e,warm} \la$ 10 cm$^{-3}$ in the warm ionized gas phase ($T \sim 10^4$ K) at the base of the NW dusty clouds entrained in the wind \citep[$R \sim 21$ kpc; aperture \#3 in Fig.\ \ref{fig:apertures} = aperture \#5 in the nomenclature of][]{Rupke2023}, corresponding to an upper limit on the mass density of the plasma $\rho_{\rm warm} \la 1 \times 10^{-23}$ g cm$^{-3}$.  Assuming pressure equilibrium between the warm and hot gas phases, the electron density in the hot gas phase at $R \sim$ 20 kpc is $n_{e,hot}$ $\la$ 0.1 cm$^{-3}$, corresponding to $\rho_{hot} \la 1 \times 10^{-25}$ g cm$^{-3}$. Unless this density is highly overestimated (which we consider unlikely; see below), we find $t_{\rm sp}$ $\ga$ 1 Myr ($a$/0.1 $\mu$m) and therefore expect that the dust grains entrained in the wind to be affected by sputtering within the outflow dynamical time scale ($\sim$ 0.1 Gyr). This is especially true for the smaller grains such as PAHs where $a$ = 0.001 $-$ 0.01 $\mu$m \citep[][]{Micelotta2010b, Richie2024}, implying $t_{\rm sp}$ $\ga$ 10$^4$ $-$ 10$^5$ yrs. 

Note that our estimate of the electron density in the hot gas phase at $R \sim 20$ kpc is an upper limit, so the sputtering time is a lower limit and could therefore, in principle, be comparable to, or even larger than, the dynamical time scale if it is severely overestimated. However, the electron density in the hot phase would have to be overestimated by a factor of $\sim$ 10$^3 - 10^4$ ($n_{e,hot} \sim 10^{-5} - 10^{-4}$ cm$^{-3}$, corresponding to $n_{e,warm} \sim 10^{-3} - 10^{-2}$ cm$^{-3}$, assuming thermal equilibrium) for the sputtering time to be comparable to the outflow dynamical time scale at $R \sim 30$ kpc ($\sim$ 0.1 Gyr); this seems unlikely for several reasons. 

First, we note that the upper error bar on the electron density measurement of \citet{Rupke2023} is rather large (1.5--2.3 dex) due to the decreasing sensitivity of the optical line ratios to densities below $\sim$ 50 cm$^{-3}$. The upper limit on the electron density of the warm-ionized gas phase at $R \sim 20$ kpc is therefore formally $n_{e,warm}$ $\la$ 10 $-$ 50 cm$^{-3}$. For comparison, the electron density of the warm-ionized wind within the central aperture of \makani\ (i.e.\ the blueshifted component of \oii\ with the large velocity dispersion within $R = 0 - 5.5$ kpc) is $\sim$ 2500 cm$^{-3}$, or $\ga$ 50--250 $\times$ larger than that of the warm-ionized wind component at $\sim$ 20 kpc. This density contrast is larger than that of the warm-ionized wind in M82 \citep[$\propto R^{-1.17}$;][]{Xu2023}, but approximately consistent with adiabatic expansion ($\propto R^{-2}$). A much lower value of $n_{e,warm}$ at $R \sim 20$ kpc would imply a very steep density gradient of the form $R^{-3}$ or $R^{-4}$, which would be unusual for galactic winds and difficult to explain physically \citep[e.g.,][]{Veilleux2005, Veilleux2020}. 

Although there are no direct estimates of the electron density in the hot gas phase of the \makani\ wind, estimates in the X-ray emitting wind fluid of M82 may be used as a guide. Using deep \chandra\ data, \citet{Lopez2020} infer a flat density profile where $n_{e,hot}$ $\sim$ 0.01 $-$ 0.03 cm$^{-3}$ at $R$ = 1 $-$ 3 kpc, values that are roughly consistent with our adopted estimate of $n_{e,hot}$ at $R \sim$ 20 kpc in \makani. Note, finally, that our estimates of $n_{e,warm}$ and $n_{e,hot}$ in \makani\ refer to the {\em inner} ($R \sim 20$ kpc) CGM, which is mass-loaded by the galactic wind. They should not be compared with the smaller values inferred from refractive Fast Radio Burst (FRB) scattering measurements \citep[$n_{e,warm} \sim 10^{-3} - 10^{-1}$ cm$^{-3}$; e.g.,][and references therein]{Mas-Ribas2025}, and used in numerical simulations \citep[$n_{e,hot} \sim 10^{-5} - 10^{-3}$ cm$^{-3}$; see references in][]{Tumlinson2017, Faucher2023}, which typically refer to the CGM at $\sim$ 100 kpc. But note that even these small values of $n_{e,warm}$ and $n_{e,hot}$ imply a thermal sputtering time scale for the smallest dust grains that is shorter than, or comparable to, the outflow dynamical time scale (0.1 Gyr). 

In summary, thermal sputtering of the small dust grains responsible for the mid-infrared emission detected with \jwst\ is expected to be significant in the dense, inner CGM of \makani. Moreover, other processes such as non-thermal (inertial) sputtering due to the non-zero velocity of the dust grains relative to the gas associated with shocks, known to be present in \makani\ \citep[][]{Rupke2023}, are expected to further contribute to the dust grain erosion \citep[e.g.,][]{Hu2019}. Again, smaller grains such as PAHs are expected to be even more susceptible to destruction by shocks \citep[e.g.,][]{Micelotta2010a}. 

Dust destruction may be avoided if the dust is shielded from these processes through cloud-wind mixing and condensation. A (dust-bearing) cool cloud may be able to survive longer than a dynamical time if the cooling time of the mixed layer between the wind and cloud material is shorter than the advection (shear) time scale  $t_{\rm shear} = r_{\rm cl} / v_{\rm w}$, where $r_{\rm cl}$ is the cloud length and $v_{\rm w}$ is the wind speed \citep[e.g.,][]{Marinacci2010, Armillotta2016, Gronke2018, Gronke2020, Li2020, Sparre2020, Kanjilal2021, Farber2022, Abruzzo2023}. 
The cloud survival criterion is $t_{\rm cool, minmix} \la 7~t_{\rm shear}$ \citep{Abruzzo2023}, where the first term is the cooling time at $T = \sqrt{T_{\rm min} T_{\rm w}}$ where $T_{\rm min}$ is the temperature between the cloud and wind temperatures, $T_{\rm cl}$ and $T_{\rm w}$, where the cooling time is minimized. In the simulations of \citet{Richie2024}, $T_{\rm min} \sim 3 \times 10^4$ K and the masses of the clouds that meet this criterion {\em grow} with time and the dust are able to survive over $\sim$ 80 $t_{\rm sp}$, which is similar to the dynamical time scale of the outflow at $R \sim$ 30 kpc. This process may therefore help explain both the presence of PAHs within this radius and the absence of significant PAH emission beyond this radius (although we cannot exclude the possibility that the absence of PAH emission at $R \ga 30$ kpc is simply due to the detection limits of the \jwst\ data). A similar argument was recently used to help explain the extent and hour-glass morphology of the far-ultraviolet O~VI 1032, 1038 \AA\ emission in \makani, which is similar to that of \oii\ \citep[][]{Ha2025}.

If this scenario is correct, one might expect the cool ($T < 10^4$ K) gas to extend on the same scale as the PAH emission, which at first seems inconsistent with the non-detection of CO (2$-$1) and \mgii\ beyond 20 kpc. However, as pointed out in Section \ref{sec:distributions}, the F1130W/CO (2$-$1), F1800W/CO (2$-$1), F1130W/\mgii, and F1800W/\mgii\ flux ratios remain relatively constant out to distances of $\sim$ 20 kpc from the center, as expected in the scenario where the PAH/CO and \mgii/CO ratios are constant. A constant PAH/CO (1$-$0) flux ratio was also observed across the inner 2 kpc of the M82 wind by \citet{Villanueva2025}. The non-detection of CO (2$-$1) and \mgii\ beyond 20 kpc in \makani\ may therefore be due to the shallower detection limits on CO (2-1) and \mgii\ in the present ALMA and KCWI data, respectively, relative to the \jwst\ detection limits on the PAH emission, rather than the true absence of cool gas at $R \ga 20$ kpc.

\begin{deluxetable*}{l c c c r r}
 \tablecaption{Measured NIRCam and MIRI Flux Densities\label{tab:fluxes}}
 \tablehead{
 \colhead{Filters} & \colhead{Main} & \colhead{$f_\nu^{\rm nuc}$} & \colhead{$f_\nu^{\rm inner-halo}$}  & \colhead{$f_\nu^{\rm CGM-E}$} & \colhead{$f_\nu^{\rm CGM-W}$}\\
\colhead{} & \colhead{Diagnostics} & \colhead{($m$Jy)} & \colhead{($\mu$Jy)} & \colhead{($\mu$Jy)} & \colhead{($\mu$Jy)}\\
 \colhead{(1)} & \colhead{(2)} & \colhead{(3)} & \colhead{(4)} & \colhead{(5)} & \colhead{(6)}}
 \startdata
N-F187N & Pa$\beta$ 1.28 & --- & --- & ---& ---\\ 
N-F480M & PAH 3.3 & 0.19 & 10.6&  $\la$ 0.4 & $\la$ 0.3\\ 
M-F770W & Continuum (5 $\mu$m) & 0.36 & 8.5& $\la$ 0.3 & $\la$ 0.3\\ 
M-F1130W & PAH 7.7 & 2.32 & 95.0& 3.7 & 3.8 \\
M-F1800W & PAH$\,$(11.3 + 12.2) & 3.24 & 78.9& 2.6 & $\la$ 0.4 \\
M-F2100W & Continuum (14 $\mu$m) & 3.62 & 2.7 & $\la$ 0.3 & $\la$ 0.3\\
M-F2550W & H$_2$ 17.03 & 5.64 & --- & --- & ---\\
 \enddata
 \tablecomments{Meaning of the columns: (1) Name of the NIRCam (N) or MIRI (M) filters. (2) Main diagnostics in the filter bandpass. (3) Flux densities in mJy measured in the unresolved nucleus (i.e. the central PSF); the statistical uncertainties on these measurement are less than 1\%. (4) Flux densities in $\mu$Jy of the dust emission in the inner halo within an annular aperture centered on the nucleus with inner and outer radii of 1\farcs78 and 3\arcsec, respectively (see Figure 6 and text for more details); the typical uncertainties on these measurements are $\pm$~0.3 $\mu$Jy. (5) $-$ (6) Flux densities in $\mu$Jy of the dust emission in the outer halo within two circular apertures centered on the brighter mid-infrared emission beyond 3\arcsec\ (see Figure \ref{fig:apertures} and text for more details); the typical uncertainties on these measurements are $\pm$~0.3 $\mu$Jy.} 
\end{deluxetable*}

\begin{figure*}
    \centering
    \includegraphics[width=0.80\textwidth]{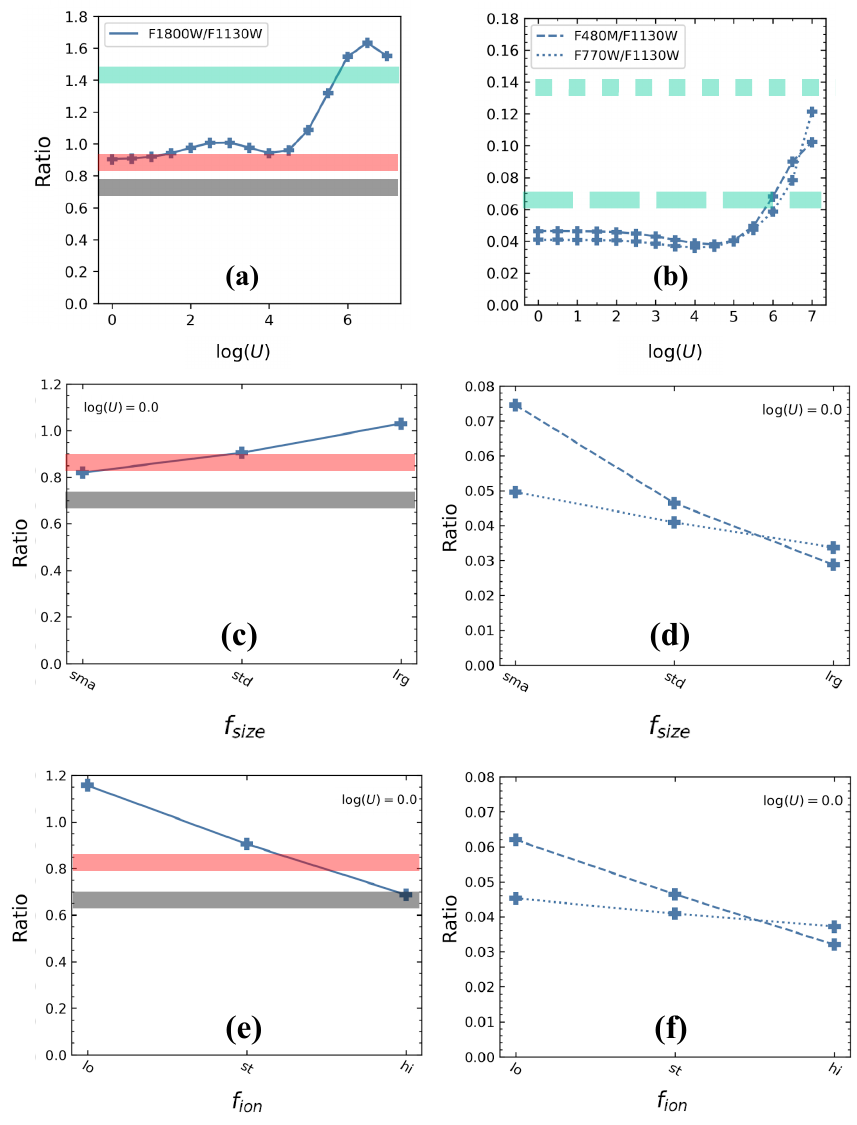}
    \caption{Measured and predicted filter flux ratios. The values of F1800W/F1130W are shown in the left panels, while F480M/F1130W and F770W/F1130W are shown in the right panels. The measured values are shown as thick, semi-transparent horizontal lines. Their colors refer to the extraction regions (nucleus = cyan, inner halo = red, outer halo = black), while the line styles refer to the filter flux ratios (F1800W/F1130W = solid line, F480M/F1130W = dashed line, F770W/F1130W = dotted line). Upper limits on the filter flux ratios are not shown in these panels (see Table \ref{tab:fluxes} and text in Section \ref{subsec:evolution}). The values predicted by the models of \citet{Draine2021} are shown as the blue curves for different values of the starlight intensities on the dust (log $U$, upper panels), PAH size distributions ($f_{\rm size}$, middle panels), and PAH ionization fractions ($f_{\rm ion}$, bottom panels).  The results in the upper panels are for standard PAH size distribution and PAH ionization fraction, while those in the middle and bottom panels are for a fixed value of log(U) = 0.0, which is appropriate for the inner and outer halos, and a standard PAH ionization fraction and PAH size distribution, respectively. See text in Section \ref{subsec:evolution} for more details.}
    \label{fig:predictions}
\end{figure*}

\subsection{Properties of the Dust in the Galactic Wind}  
\label{subsec:evolution}

The physical state of dust grains may be assessed from the relative strengths of the PAH features and surrounding continuum \citep[e.g.,][]{Draine2021, Rigopoulou2021, Rigopoulou2024}. If the processes of cloud-wind mixing and condensation are indeed important in the CGM of \makani, the PAH flux ratios may be expected to vary radially as the clouds are entrained in the outflow.

In Figure \ref{fig:predictions}, we show the predicted F1800W/F1130W, F770W/F1130W, and F480M/F1130W flux ratios for the \citet{Draine2021} library of PAH + dust emission spectra, redshifted to $z$ = 0.459, for different starlight intensities ($U$), PAH size distributions ($f_{\rm size}$), and PAH ionization fractions ($f_{\rm ion}$). For each model, the predicted fluxes were calculated using the transmission curves of the NIRCam and MIRI filters (Fig.\ \ref{fig:filter_throughput_makani}) without applying any continuum subtraction. $U$ is defined as the normalized energy density such that identical values of $U$ yield the same temperature of dust grains from heating. The value $U$ = 1 refers to the heating that would be produced by the starlight spectrum of the solar neighborhood. The $f_{\rm size}$ parameter describes the distribution of grain sizes comprising the PAH population, and is described by three parameter values---small (sma), standard (std), and large (lrg). Finally, the $f_{\rm ion}$ parameter describes the fraction of ionized PAH relative to the total population. The \citet{Draine2021} models explore three values of $f_{ion}$: low (lo), standard (st), and high (hi). 
For this figure, we used the \citet{Draine2021} models with the starlight spectrum predicted from the single-age stellar population models of \citet{Bruzual2003} for solar metallicity ($Z=0.02$) and a stellar age of 10 Myr, representative of local $L^*$ star-forming galaxies. In Shockley et al. (2025, in prep.), we show that the flux ratios predicted for other stellar models with either younger stellar age (3 Myr) or lower metallicity ($Z = 0.0004$ = 2\% solar) lie within 5\% of each other. 

All three filter flux ratios are sensitive to the starlight intensity, especially for values of log $U$ above 4. F1800W/F1130W and F480M/F1130W also provide some constraints on the PAH size distributions and PAH ionization fractions, especially for small values of log $U$. We compare these predictions with the measured values of these ratios listed in Table \ref{tab:fluxes} for the nucleus (PSF), inner halo (purple annulus at $R$ $\sim$ 10 $-$ 20 kpc in Fig.\ \ref{fig:apertures}), and outer halo (red circular apertures in Fig.\ \ref{fig:apertures} centered on the NW circumgalactic cloud complex at $R$ $\sim$ 20 $-$ 35 kpc). In the nucleus, the measured high values of F1800W/F1130W ($\sim$ 1.4), F480M/F1130W ($\sim$ 0.07), and F770W/F1130W ($\sim$ 0.14) all point to large starlight intensities, PAH sizes that are larger than the standard distribution, and small ionization fractions. 

In contrast, the smaller values of F1800W/F1130W in both the inner halo ($\sim$ 0.83) and outer halo ($\sim$ 0.70) can only be reproduced if the starlight intensity is low (log $U$ $\la$ 0), the PAH sizes are smaller than the standard distribution, and the ionization fractions are elevated relative to the standard ionization distribution of \citet{Draine2021}. The upper limits of $\la$ 0.1 on the F480M/F1130W flux ratios in both the inner and outer halo regions are consistent with these conclusions, but are too high to provide additional physically meaningful constraints. The upper limits on the F770W/F1130W flux ratios in these regions ($\la$ 0.1) also favor small log $U$ values, but do not constrain the PAH sizes and ionization fractions.

The results of \makani\ seem qualitatively different from the \spi\ results in M82. The PAH 11.3/7.7 ratios in the northern wind of M82 {\em increase} with increasing distance from the midplane, particularly beyond 2 kpc, indicating that the PAH emitters in the wind are more neutral than in the starburst disk \citep[][]{Beirao2015}. \footnote{Note that the recent \jwst\ analysis of \citet[][]{Villanueva2025} covers the region within 2 kpc from the center of M82 so it cannot confirm the results of \citet{Beirao2015}.} Cold clouds in the wind of M82 may not only survive the erosive processes associated with the acceleration by a warm, fast wind, but grow in mass as they flow out, due to radiative cooling in the mixed gas \citep{Gronke2018, Gronke2020, Schneider2020, Banda-Barragan2021} or thermal instabilities in the hot fluid \citep{Thompson2016}. On the other hand, in \makani, the smaller F1800W/F1130W flux ratios in the inner and outer halos at $r$ $\sim$ 15 kpc and  $\sim 35$ kpc, respectively,  relative to the value in the nucleus indicates a radially decreasing PAH (11.3 + 12.2)/7.7 ratio profile. This is due in large part to the radially decreasing starlight intensity. But, as discussed above, the actual F1800W/F1130W flux ratios in the inner and outer halos also imply a reduction of the average PAH sizes and increase in PAH ionization with increasing radial distances, both pointing to PAH destruction on the journey to the outer halo \citep[e.g.,][]{Narayanan2023}.  The outflow timescale ($R/v$) of \makani\ is $\sim$ 10 $-$ 100 $\times$ that of M82, so we may be seeing long-term dust evolution in \makani\ which is simply not probed by the M82 \spi\ data. This erosion of the dust grains in \makani\ may also coincide with the dissolution of the outflowing cool clouds into the warm ionized phase, as predicted by theory \citep[e.g.,][]{Scannapieco2015, Ferrara2016, Decataldo2017}. However, we cannot formally rule out the possibility that both the dust {\em and} cool gas survive out to $\sim$ 35 kpc, and perhaps even beyond, given the detection limits on  the cool gas (warm dust) beyond $\sim$ 20 (35) kpc in the existing ALMA and KCWI (\jwst) data (Section \ref{subsec:survival}).

\section{Conclusions}
\label{sec:conclusions}

We report the results of our analysis of deep NIRCam and MIRI images centered on the PAH 3.3, 7.7, and (11.3 + 12.2) features and adjacent continuum regions in the \makani\ galaxy, a compact post-starburst galaxy ($r_e$ = 2.3 kpc) at $z$ = 0.459 with a warm-ionized wind that spans 100 kpc. The main conclusions are as follows:

\begin{itemize}
    \item[$\bullet$] A careful removal of the bright central source in this object using \stpsf\ reveals PAH 7.7 and (11.3 + 12.2) emission that extends to $\sim$ 35 kpc and $\sim$30 kpc, respectively, well beyond the \mgii\ and CO 2$-$1 line emission associated with the more recent (7 Myr) starburst episode, but not as far as the outer \oii\ line emission produced by the older (0.5 Gyr) episode. 
    
    \item[$\bullet$] The discovery of PAHs in the warm ionized wind indicates that they survive the long journey ($R/v$ $\sim$ 10$^8$ yrs) to the CGM. Dust destruction due to thermal sputtering and the many other processes taking place in the harsh environment of the galactic wind may be avoided if the dust is shielded through cloud-wind mixing and condensation.  The roughly constant PAH-to-CO and PAH-to-\mgii\ flux ratios with radial distances from the center are consistent with this scenario. Cool gas may be present beyond 20 kpc, but is beyond the detection limits of the existing ALMA and KCWI data. Similarly, the detection limits of our \jwst\ data may prevent us from detecting warm-dust emission beyond $\sim$ 35 kpc.
    
    \item[$\bullet$] The F1800W/F1130W flux ratio, tracer of the PAH (11.3 + 12.2)/7.7 ratio, is smaller in the inner halo at $R$ = 10 $-$ 20 kpc and outer halo at $R \sim$ 20 $-$ 35 kpc than in the unresolved nuclear emission, indicating that the PAH emitters in the inner and outer halos are not only more weakly illuminated than in the nucleus, which is expected, but also smaller and more ionized. This is the first evidence for dust evolution on a outflow dynamical time scale of $\sim$ 10$^8$ yrs. This result is qualitatively different from that found in M82, where the PAH (11.3 + 12.2)/7.7 ratio increases over a dynamical time scale of $\sim$ 10$^6$ yrs. The measured upper limits on the F480M/F1130W and F770W/F1130W flux ratios, tracers of the PAH 3.3/7.7 and 5 $\mu$m continuum/PAH 7.7 ratios in these regions, are consistent with these radial trends. 

    \item[$\bullet$] Overall, these results provide the strongest evidence so far that the ejected dust of galactic winds may survive the long journey to the CGM, but erodes along the way.
\end{itemize}

\begin{acknowledgments}

We thank the anonymous referee for constructive comments that helped improve this paper. This work is based [in part] on observations made with the NASA/ESA/CSA James Webb Space Telescope.   These observations are associated with program \#1865. The \jwst\ data presented in this article were obtained from the Mikulski Archive for Space Telescopes (MAST) at the Space Telescope Science Institute. The specific observations analyzed can be accessed via \dataset[doi: 10.17909/1r37-nb46]{https://doi.org/10.17909/1r37-nb46}. Support for program \#1865 was provided by NASA through a grant from the Space Telescope Science Institute, which is operated by the Association of Universities for Research in Astronomy, Inc., under NASA contract NAS 5-03127.

\end{acknowledgments}

%

\vspace{5mm}

\begin{contribution}
SV conceived and led the implementation of the \jwst\ program, helped with the data analysis, led the interpretation of the data, and wrote the manuscript. 
SDS produced the figures and led the comparison of the measured filter flux ratios with the predicted values of the PAH + dust models. MM carried out the \stpsf\ analysis and edited the \jwst\ proposal and manuscript, particularly Section \ref{sec:data-reduction-analysis}. DSNR helped with the analysis and figures and edited the \jwst\ proposal and manuscript. CAT and HC provided the nuclear spectrum shown in Fig.\ \ref{fig:filter_throughput_makani} and edited the \jwst\ proposal and manuscript. ALC, AMDS, JEG, BCH, JM, GHR, and PHS helped improve the \jwst\ proposal and manuscript. 


\end{contribution}

\facilities{JWST(MIRI, NIRCam)}


\software{astropy \citep{Astropy2013, Astropy2018},  \stpsf\ \citep[][]{Perrin2014}}




\bibliography{makani}{}
\bibliographystyle{aasjournalv7}



\end{document}